\documentclass[journal,twoside,web]{ieeecolor}
\usepackage{generic}
\usepackage{amsmath,amssymb,amsfonts}
\usepackage{graphicx}
\usepackage{textcomp}
\usepackage{epstopdf}
\usepackage{siunitx}
\usepackage{amssymb}
\usepackage{subfig}
\usepackage{cite}
\usepackage{ mathrsfs }
\def\BibTeX{{\rm B\kern-.05em{\sc i\kern-.025em b}\kern-.08em
		T\kern-.1667em\lower.7ex\hbox{E}\kern-.125emX}}
\begin{document}
	\title{Charge-based Modeling of Ultra Narrow \\Cylindrical Nanowire FETs} 
	\author{Danial Shafizade, Majid Shalchian, and Farzan Jazaeri
		\vspace{-0.6cm}
		\thanks{
			Danial Shafizade and Majid Shalchian are with the Electrical Engineering Department, Amirkabir University of Technology, 424 Hafez Ave, Tehran, Iran, 15875-4413. Farzan Jazaeri is with the Integrated Circuits Laboratory (ICLAB) of the Ecole Polytechnique F\'{e}d\'{e}rale de Lausanne, Switzerland.}}
	\maketitle
	\begin{abstract}
		This brief proposes an analytical approach to model the dc electrical behavior of extremely narrow cylindrical junctionless nanowire field-effect transistor (JL-NW-FET). The model includes explicit expressions, taking into account the first order perturbation theory for calculating eigenstates and corresponding wave functions obtained by the Schr\"{o}dinger equation in the cylindrical coordinate. Assessment of the proposed model with technology computer-aided design (TCAD) simulations and measurement results confirms its validity for all regions of operation. This represents an essential step toward the analysis of circuits mainly biosensors based on junctionless nanowire transistors.
	\end{abstract}
	\section{Introduction}
\IEEEPARstart{J}{unctionless} (JL) Silicon nanowire FETs with heavily doped channel, typically in the range of $10^{19} $ cm$^{-3}$ to $10^{20} $ cm$^{-3}$, have been the subject of intensive research both in the fields of fabrication process and modeling. Due to its extremely interesting morphological and functional bio-compatible for sensing applications \cite{singh2019vertical,smith2019surface}, silicon nanowire has been broadly used for the label-free detection in chemical and biological experiments\cite{Nature,lee2019label}.  Given the advantage of using junctionless silicon nanowires in a wide range of applications, especially for sensing applications, several compact models have been proposed so far \cite{6403540, 8374091}. Relying on the Poisson-Boltzmann equations, the charge-based models were developed to model junctionless double-gate and nanowire FETs in \cite{5872019, Book, 6879262}. However, those models overestimate the drain to source current and free carrier charge density for the channel thicknesses less than $10$ nm due to the charge quantization into discrete sub-bands, which was not taken into account in classical models. To accurately capture the quantization effects, we recently proposed a charge-based model, including quantum confinement for ultra-thin junctionless double-gate FETs and GAA Junctionless nanowire \cite{8424086} and \cite{D.Shafizade2019.9}. The proposed approach relies on the Schr\"{o}dinger equation with the $ 1^{st} $ order perturbation theory and Fermi-Dirac statistics. 
In this context, extending the proposed approach, we derive an explicit model for extremely narrow cylindrical junctionless nanowire FET which has easier fabrication and higher surface to volume ratio compared to rectangular cross-section nanowire, which are important features for nano-scaled sensor structures. cylindrical device also exhibit uniform distribution of potential and electric field around its surface without electric field crowding near the corners, making this structure a candidate for sensing application. Thus, taking into account the stated advantages, we concentrate on the derivation of a simple analytical model for extremely narrow cylindrical junctionless nanowire FET for biosensing applications.
\section{Electrostatics in Ultrathin Junctionless Cylindrical Nanowire FETs}
We consider an \textit{n}-type junctionless cylindrical nanowire with diameter $D$ and radius $R$. Fig. 1(a) shows a 3D  schematic view of a cylindrical junctionless nanowire. When the gate to channel voltage is below the flat-band potential, we may assume full depletion approximation and neglect free carrier charge density assuming parabolic approximation for the potential profile across the channel as a solution of the Poisson equation. Hence, the potential distribution across the channel can be expressed simply as:
\begin{equation} \label{1}
\psi (r) = ({\psi _s} - {\psi _c})(\frac{4r^2}{D^2}-1) + {\psi _s},
\end{equation}
Where $\psi _s$ and $\psi _c$ are the electrostatic potentials at the center and surface of the cylindrical channel. On the contrary, by increasing the gate voltage above the threshold, the density of electrons in the few lowest subbands increases rapidly according to Fermi-Dirac statistics in a system with 2D confinement \cite{Numata_2010}. Fig. 2(a) represents the potential profile across the channel, which compares TCAD simulation results with the proposed model for $D = 5$ nm, from depletion to accumulation mode. Physical parameters for cylindrical JL nanowire FET used in the TCAD simulations and model are listed in Table 1. 
This confirms the validity of parabolic approximation for both depletion and accumulation modes, by properly including the impact of carrier charges in ($\psi_{s} - \psi_c$) term. As seen, the second derivative of the potential profile versus $ r $, which is proportional to the total charge density, becomes positive/negative when the device enters into accumulation/depletion modes of operation. While the channel is neutral and $(Q_{sc}=0)$, the values of surface and center potentials become equal, i.e. flat-band condition (see Fig. 1(b)). The electric field at the channel-oxide interface, $E_s$, can be obtained using (1) and is linked to the overall space charge density. Imposing the boundary conditions at the perimeter of the channel, the electric field can be given as a function of center and surface potentials by 
	\begin{equation} \label{2}
	{Q_{sc}} = \pi D{\varepsilon _{si}}\frac{{d\psi (r)}}{{dr}}\bigg| _{r = R} =- 4\pi {\varepsilon _{si}}\Delta V.
	\end{equation}
	\begin{figure}[!t]
		\centerline{\includegraphics[height=8cm, width=8cm]{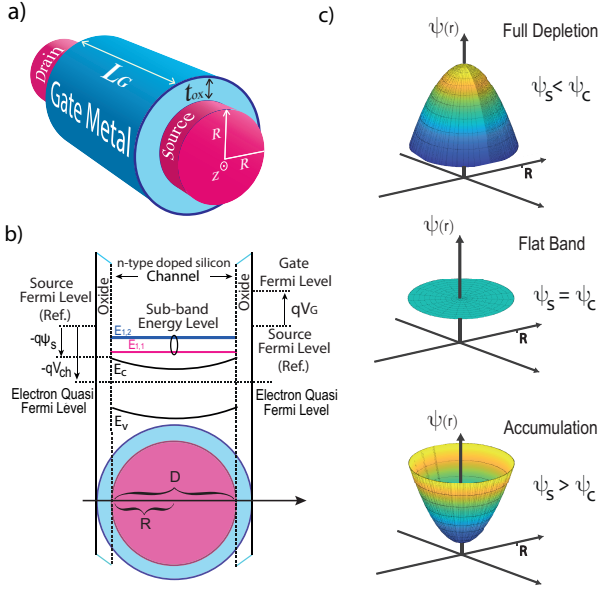}}
		\caption{(a) 3D view of Cylindrical Junctionless Nanowire. (b) energy band-diagram across the diameter of the circular cross section. (c) 3D Potential distribution across the channel cross section, for full depletion, flat band and accumulation mode.}
		\label{fig1}
		\vspace{-0.5cm}
	\end{figure}
where $\Delta V = \psi_{s} - \psi_c$. For any channel cross section with radius R, one may write the Gauss law in circular coordinate:
\begin{equation} \label{3}
	Q_{sc}=-2\pi DC_{ox}(V_{GS}-\Delta \phi_{ms}-\psi_{s}).
\end{equation}
$C_{ox}$ is the gate dielectric capacitance per unit area given by
\begin{equation} \label{4}
C_{ox}=\varepsilon_{ox}\left[R\times \ln\left(1+t_{ox}/R\right)\right]^{-1}.
\end{equation}
Combining (2) and (3) gives a relationship between surface and center potentials for a given gate voltage:
\begin{equation} \label{5}
	{\psi _s} = {V_{GS}} - \Delta {\phi _{ms}} - \frac{{{2\varepsilon _{si}}}}{{D{C_{ox}}}}\Delta V.
\end{equation}
On the other hand, the total charge density per unit length is the sum of the mobile charge density, $Q_m$, and fixed-charge density, $Q_{fix}=q \pi N_{D} R^2$, given by
	\begin{align}
	{{Q}_{sc}}={{Q}_{m}}+q \pi N_{D} R^2.
	\end{align}	
Here mobile charge calculate by assuming 2D confinement of electron in the channel for ($D<8$ nm) and relying on Fermi Dirac integral in order of $-\frac{1}{2}$ and 1D density of state \cite{commonJazaeri2013, doi:10.1063/1.4759275} we propose to estimate the mobile charge density, $Q_m$, by
	\begin{align}
	{{Q}_{m}}=-\sum\limits_{k=1}^{2}{qDo{{S}_{k}}{{F}_{-\frac{1}{2}}}}\left( -\frac{\eta_{k} }{qv_t} \right),
	\end{align}
	where $\eta_{k}=E_{k,n=1}^{T}-{q{\psi}_{s}}+{q{V}_{ch}}$ is the difference between the first sub-band eigenstate energy level and electron quasi-Fermi level  $(qV_{ch})$. As shown in Fig. 1(c) electron quasi-Fermi level varies from $ 0 $ at the source to $-qV_{DS}$ at the drain terminal. Moreover, $DoS_k$ represents 1D density of state. we further assume that only two degenerate valleys ($k=1,2$) of the first sub-band ($n=1$) contribute to $Q_m$.
	\section{Eigenstate Energy Levels in Cylindrical Coordinate and Electron Charge Density}
	 As stated in \cite{8424086} Solving the Schr\"{o}dinger equation, for an ideal circular infinite quantum well around the channel at Si-SiO$_2$ interface, eigenstate energy levels and electron wave-functions in discrete sub-bands are given by
\begin{equation}
	E_{k,n} = \frac{{{\hbar ^2}{\alpha _n}^2}}{{2{m_k^*}{R^2}}},
\end{equation}
	\begin{equation} 
   \Psi(r) ={\frac{\sqrt{2}}{RJ_1(\alpha_n)}J_{o}(\frac{ \alpha _n r}{R})}.
   \end{equation}
	\vspace{-1.1cm}
	\begin{center}
		\begin{table}
			\caption{Physical parameters of cylindrical JL nanowire FET used in the TCAD simulations and model derivation}
			\renewcommand{\arraystretch}{1.5}
			\begin{tabular}{p{4cm}p{1.3cm}p{2.3cm}}
				\hline
				\textbf{Parameter} &\textbf{Symbol}&\textbf{Value}  \\
				\hline
				\hline
				\textbf{Channel Doping} &${N}_{D}  $&$10^{19} $ cm$ ^{-3} $  \\
				\textbf{Channel Diameter}  &$D $& $4-6$ nm  \\
				\textbf{Oxide Thickness}  &${t}_{ox} $& $1$ nm  \\
				\textbf{Channel Width}&${L}_{G} $&$4$ $\mu$m  \\
				\textbf{Permittivity in Vacuum }&$\varepsilon_{o}$ &$8.85{\times}10^{-12}$ F/m  \\
				\textbf{Silicon Permittivity}&$\varepsilon_{si}$&$11.7\varepsilon_{o}$   \\
				\textbf{Silicon Oxide Permittivity}&$\varepsilon_{ox}$&$3.9\varepsilon_{o}$   \\
				\textbf{longitudinal  Effective mass }&$m_{r1}$ &$0.19m_{o}$  \\
				\textbf{Transverse Effective mass}&$m_{r2}$ &$0.315m_{o}$  \\
				\textbf{Silicon Band Gap}&$E_{g}$ &$1.12$ eV  \\
				\textbf{Gate Work Function}&$\Delta\phi_{m}$ &$4.8$ eV  \\
				\textbf{Conduction Band effective DoS}&$N_{c}$ & $2.8{\times}10^{19}$ cm$^{-3}$  \\
				\textbf{Valence Band Effective DoS}&$N_{v}$ &$1.04{\times}10^{19}$ cm$^{-3}$  \\
				\textbf{Silicon Intrinsic Carrier Density}&$n_{i}$ &$1.45{\times}10^{10}$ cm$^{-3}$    \\
				\textbf{Temperature}&$T$ &$300^{\circ}$ K \\
				\textbf{Thermal Voltage}&$v_{t}$ & $0.025$ V  \\
				\textbf{First zero of $J_0(r)$}&$\alpha_{1}$ & $2.4048$ \\
			     \textbf{Second zero of $J_0(r)$}&$\alpha_{2}$ & $5.5201$  \\
				\hline
			\end{tabular}
		\end{table}
	\end{center}
Here, $J_0$ and $J_1$ are Bessel functions of the first kind and $\alpha_n$ is $n^{th}$ zero of $J_0(r)$. The terms of $\alpha_{1}=2.4048$ and $\alpha_{2}=5.5201$ are used to calculate the first and the second eigenstate energy levels in the absence of electrostatic potential perturbation respectively. The term of $m_k^*$ corresponds to the electron effective mass inside the channel. In this study, the effective masses are taken as $m_{r1}$ $ = 0.19 $$m_o$ for the first valley and $m_{r2}$ = $ 0.315 $$m_o$ for the second valley in a $<110>$-oriented Si channel \cite{6151818}. The first order perturbation theory is used to calculate the impact of electrostatic potential on sub-band energies:
	\begin{equation} \label{7}
	E_{n}^{p}=q\int\limits_{0}^{R}\left\{{\left[ \Psi (r) \right]^{*}{\psi }(r){{\left[ \Psi (r) \right]}}}\right\}rdr.
	\end{equation}
	Substituting $\psi(r)$ from (1) and $\Psi (r)$ from (9) into (10) leads to an analytical expression for the first order perturbation energy in the cylindrical coordinate:
	\begin{equation} \label{7}
	\begin{split}
	E_{n}^p = q\frac{2\Delta V}{3}\bigg(\frac{\alpha_n+1}{\alpha_n}\bigg).
	\end{split}
	\end{equation}
	Total sub-band energy which results from geometrical and electrical confinements is the sum of energies obtained in (8) and (11):
	\begin{equation} \label{8}
	E_{k,n}^T = E_{k,n} + E_{n}^p.
	\end{equation}
  Fig. 2(b) and (c) demonstrate $E_{k,n}^T-q\psi_{s}$ for two first sub-bands and degenerate valleys measured from electron energy at the channel surface as a reference. Although simulation results show a good agreement compare to model for both sub-bands and degenerate valleys but only the first sub-band considered in model calculations, which reduce model computation. The occupation percent of electrical charge in the two lowest sub-bands indicates that the error in the charge density is less than $1.2\%$ for $6$ nm channel diameter when we neglect contributions of higher subbands ($n>1$) in charge density calculations. However, to calculate the charge density more preciously, additional sub-band should be included. Fig. 2(d) and (e) show the mobile charge density concerning the effective gate voltage for different channel diameters, i.e., $D=4, 5, 6, 8$ nm calculated from the proposed model and validated by TCAD simulations results. In what follows, we discuss drain-current derivation in which we use (7) and (12) to calculate channel charge density.

\section{Drain Current Derivation in Cylindrical JL-NW-FET}
As done in \cite{D.Shafizade2019.9}, relying on the  drift-diffusion transportation mechanism, the total drain current for both depletion and accumulation regions can be obtained by
	\begin{align}
	{{I}_{DS}}=-\frac{\mu  q}{L_G}\int\limits_{{{\eta}_{S}}}^{{{\eta}_{D}}}{{{Q}_{m}}}\frac{dV_{ch}}{d\eta}d\eta.
	\end{align}
Here, the term $\mu$ is the free carrier mobility, ${\eta}_{S}=E_{k,n}^{T}-q{\psi}_{s}$ and ${\eta}_{D}=E_{k,n}^{T}+q{V}_{DS}-q{\psi}_{s}$ define as the differences between the first sub-band and quasi-Fermi levels at the source and drain sides of the channel. Differentiating $\eta$ with respect to the $V_{ch}$ we get:
	\begin{equation}
	\frac{d\eta}{dV_{ch}}=q\left[ 1+\frac{d{{\psi }_{s}}}{d{{V}_{ch}}}\left( \frac{dE_{k,n}^{T}}{d{{\psi }_{s}}}-1 \right) \right].    
	\end{equation}
	Using (5), $d{\psi }_{s}/d{V}_{ch}$ can be expressed as
	\begin{equation}
	\frac{d{{\psi }_{s}}}{d{V}_{ch}}=\frac{1}{{2\pi DC_{ox}}}\frac{d{{Q}_{m}}}{d{V}_{ch}}.
	\end{equation}
\begin{figure*}[t!]
	\vspace*{-0.7cm}
	\centering
	\includegraphics[width=0.84\textwidth]{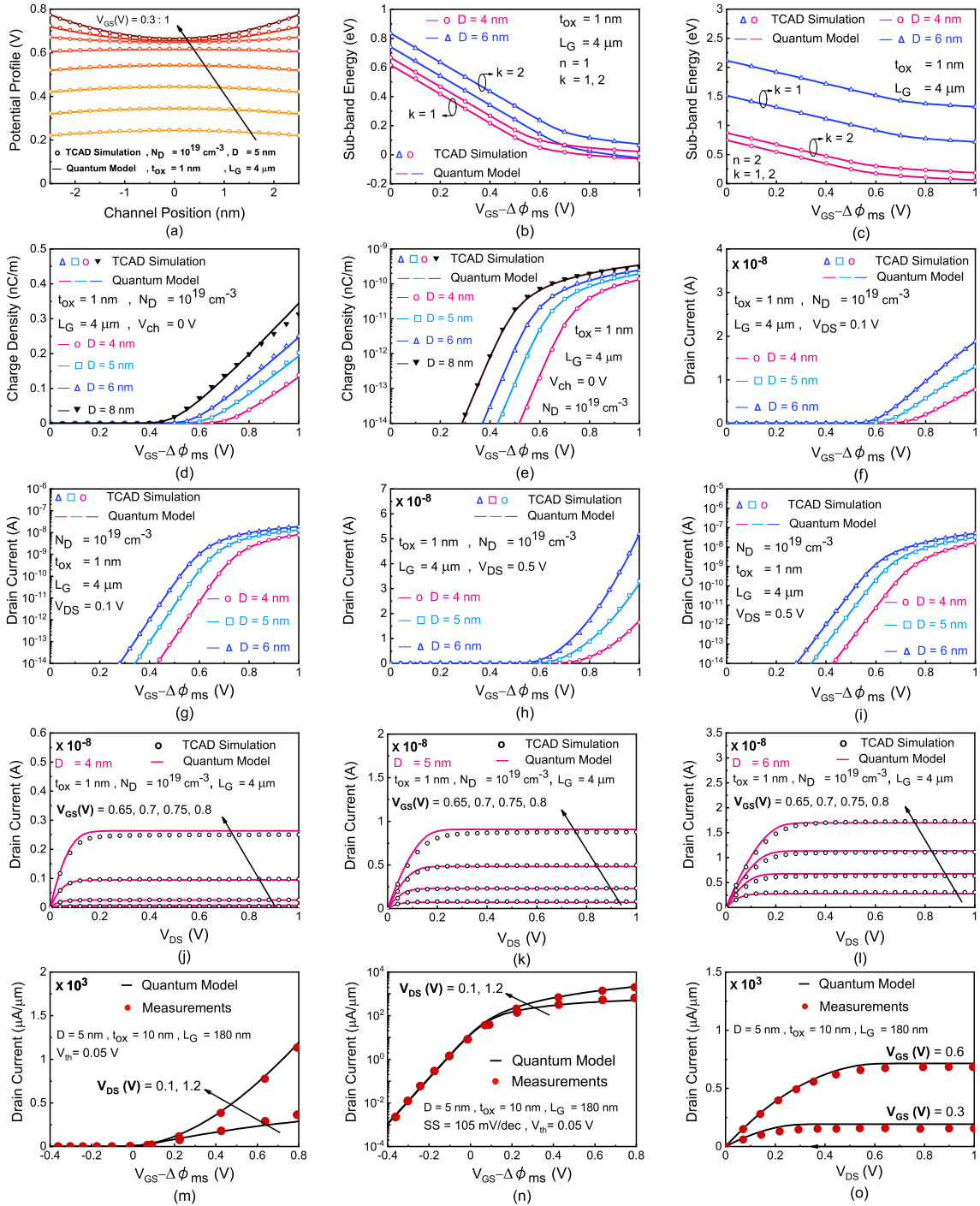}
	\caption{(a) Potential distribution across the $5 nm$ channel diameter for different values of the gate potential. Solid lines: proposed model. Symbols: TCAD simulation. (b) And (c) shows respectively first sub-band energy ($n=1$) and second sub-band energy ($n=2$) from the reference of electron quasi-fermi level at surface of the channel as a function of gate voltage for channel diameter of $4$ and $6$ nm for two degenerate valley calculated from analytical model (Solid lines) and verified by TCAD simulation (symbols). Mobile charge density with respect to the effective gate to source voltage for different channel diameters of $4, 5, 6$ and $8$ nm (d) linear scale and (e) semi-log scales. (f) and (g) represented drain current versus effective gate voltage for $V_{DS}=0.1$ V for linear and logarithmic scale, respectively. (h) Linear and (i) logarithmic scale of drain current with respect to the gate source voltage for $V_{DS}=0.5$ V. (j), (k) and (l) shows drain current versus drain voltage for several values of $V_{GS}$ and channel diameter of $4$, $5$ and $6$ nm respectively. Solid lines: proposed model. Symbols: TCAD simulation. (m) linear and (n) logarithmic scale of normalized drain current versus gate voltage for $V_{DS}=0.1$ V and $V_{DS}=1.2$ V from the analytical model (solid and dash lines) and experimental measurements (symbols). (o) Show normalized drain current as a function of drain voltage for $V_{GS}=0.3$ V and $V_{GS}=0.6$ V. Analytical model with $f=1$ (solid lines) and experimental measurements (symbols). Measurements are taken from \cite{1626464}.}
	\label{fig 2}
	\vspace*{-0.7cm}
\end{figure*}
	Next, using (11) and (5) and applying the chain rule  $d{E_{k,n}^{T}}/d\psi_{s} = d{E_{k,n}^{T}}/d\Delta V \times d\Delta V/d\psi_{s} $, we obtain: 
	\begin{equation}
	\begin{split}
	\frac{d{E_{k,\alpha_n}^{T}}}{d{{\psi }_s}}\!=\Bigg(\!\frac{\alpha_n\!+\!1}{\alpha_n}\!\Bigg)\Bigg(\!\frac{C_{ox}D }{3\varepsilon_{si}}\!\Bigg)=\!a\!
	\end{split}
	\end{equation}
	Here, $a$ is a key parameter which depends on the channel diameter and gate oxide thickness. Now we can write ${dV}_{ch}/{d\eta}$ based on (14), (15) and (16) as follows:
	\begin{equation}
	\frac{d{V}_{ch}}{d\eta}=\left[ \frac{1}{q}-\beta \frac{dQ_m}{d\eta } \right].
	\end{equation}
	where $\beta= (a-1)/2\pi DC_{ox}$.
	Substituting (17) into (13) leads to:
	\begin{equation}
	{{I}_{DS}}=\frac{\mu}{L_G}\int\limits_{{{\eta}_{S}}}^{{{\eta}_{D}}}{{{Q}_{m}}}d\eta -\frac{\beta\mu q}{L_G}\int\limits_{{{\eta}_{S}}}^{{{\eta}_{D}}}{{{Q}_{m}}}dQ_m.
	\end{equation}
	Replacing $Q_m$ from (7) in (18), and after integrating, we obtain an analytical expression for the drain current:
	\begin{equation}
	\begin{split}
	{{I}_{DS}}=&\frac{q \mu v_t}{{{L}_{G}}}\sum\limits_{k=1}^{2}{Do{{S}_{k}}}\left[ {{F}_{\frac{1}{2}}}(-\frac{\eta }{qv_t}) \right]_{{{\eta }_{S}}}^{{{\eta }_{D}}}\\
	&-\frac{\mu\beta }{2{{L}_{G}}}\sum\limits_{k=1}^{2}{{{q}^{2}}DoS_{k}^{2}}\left[ \Bigg({{F}_{-\frac{1}{2}}}(-\frac{\eta }{qv_t}) \Bigg)^{2}\right]_{{{\eta }_{S}}}^{{{\eta }_{D}}}.
	\end{split}
	\end{equation}
	\section{Results and Discussion}
Fig. 2(f) and (h) show the drain current versus the effective gate voltage at low ($V_{DS} = 0.1 V$) and high ($V_{DS} = 0.5 V$) drain potentials for linear and saturation modes of operation. Fig. 2(j), (k), and (l) show the drain current with respect to $V_{DS}$ for $D=4-6$ nm respectively for several $V_{GS}$ above the flat-band condition. Additionally, fig. 2(m), (n), and (o) show the drain current comparison of model results and experimental data for cylindrical junctionless nanowire with $D= 5$ nm, $t_{ox}=10$ nm and $L_{G}=180$ nm \cite{1626464}. To compare our model which is based on long channel approximation with experimental data with $180$ nm channel length, we introduced a correction factor $f=1.6$ in such a way that $v_{t}$ in (7) will now be replaced by $fv_{t}$ to recover the mismatch in the subthreshold region for drain current (see Fig.2 (n)). It worth highlighting that for long channel length devices, the suggested model gives an accurate prediction of the carrier charge density without any correction factor. 

A self-consistent coupled Schrödinger-Poisson model has been used in TCAD simulations for mobile charge density and band state energy derivation. For the current calculation, we used the Bohm Quantum Potential model in our simulation. TCAD assures that there is a close agreement between Bohm Quantum Potential and the results of Schr\"{o}dinger-Poisson model calculations for any given class of device. The result predicted by the model shows excellent agreement with TCAD simulations, in both linear and saturation regions. TCAD simulation results verify that the proposed model works well for both depletion and accumulation regions for the device with an extremely narrow channel size ($D=4-6$ nm). 

	Before concluding, we would like to say few words about the limitations of this model. As simulation results indicated, the model demonstrates excellent accuracy for $D= 4-6$ nm. However, as shown in Fig.2(d), for channel diameter larger than $6$nm the model predicted charge density deviates from simulation results. Indeed, by increasing channel diameter, the potential distribution in the channel does not follow the parabolic approximation anymore. As a result, first-order perturbation theory (11) can not correctly predict discrete sub-bands' energy and higher order interactions should be taken into account.  Moreover, the energy distance between sub-bands decreases and therefore the charge in several subbands participate in the conduction process. Another important consideration is variability due to random fluctuations in ultra-thin devices with high doping concentration. A capillary channel with high doping concentration caused a random position of a few impurity atoms in the channel, which has a critical impact on model validity due to device to device variations. This issue is critical for VLSI applications, however as discussed in \cite{D.Shafizade2019.9}, it is not a major concern as long as we focused on biosensing applications.

	\section{Conclusion}
In this paper, we developed a simple analytical model for 2D electrostatic potential distribution and charge density of extremely narrow cylindrical nanowire FETs. Takes in to account cylindrical coordination and first-order correction for the confined energies analytical solution developed for discrete energy sub bands. Moreover, by using drift-diffusion method we derived simple analytical expression for drain current. The validity of the model verified by TCAD simulations and also experimental data for a device with channel diameter in the range of $4$ to $6$ nm. Results evaluation confirms that the proposed model can capture precisely the dc behavior of such device for all regions of operation. In addition, this approach might also be used to derive trans-capacitance model for ac analysis with cylindrical junctionless nanowire devices\cite{JAZAERI201434, 6642051}. 
 
\bibliographystyle{IEEEtran}

\bibliography{IEEEabrv,biblio}
\vspace{-3cm}

\end{document}